\documentclass[aps,11pt,tightenlines,superscriptaddress]{revtex4-1}
\usepackage{graphics}
\usepackage{graphicx}
\usepackage{amsmath}

\begin{document}

\title{Penetration of ELF Currents and Electromagnetic Fields into the Off-Equatorial E-Region of the Earth's Ionosphere}


\author{Neeraj Jain}
\affiliation{Department of Astronomy, University of Maryland, College Park, MD, USA.}

\author{B. Eliasson}
\affiliation{Theoretische Physik IV, Ruhr-Universit\"{a}t Bochum, Bochum, Germany.}

\author{A. S. Sharma}
\affiliation{Department of Astronomy, University of Maryland, College Park, MD, USA.}

\author{K. Papadopoulos}
\affiliation{Department of Astronomy, University of Maryland, College Park, MD, USA.}





\begin{abstract}
The generation  of ELF (of the order of  10 Hz) currents and electromagnetic fields in the off-equatorial E-region (90-120 km) of the Earth's ionosphere  and their subsequent penetration into the deeper ionospheric layers is studied theoretically and numerically.  These ELF currents and fields are generated 
by the interaction of an electromagnetic pulse with the E-region at its lower boundary located at 90 km above the Earth's surface.
The wave penetration (with a typical wavelength of the order of 10 km) of the generated ELF currents and fields into the deeper ionospheric layers up to 120 km takes place due to the dominance of the Hall conductivity over the Pederson conductivity in the region between 90-120 km and penetration becomes diffusive above  120 km. 
During night time, the increase in the wave speed  due to the reduced conductivities leads to the deeper penetration.  As the angle between Earth's magnetic field and horizontal is increased (going away from the equator), the  currents and fields penetrate deeper into the ionospheric layers with increased wavelength, the magnitudes of horizontal (east-west) and vertical currents decrease near the boundary and the vertical electric field decreases drastically. The horizontal (east-west) current integrated along vertical  is fitted with a current distribution which can be replaced by a line current raised above its actual height by the half width of the current distribution for the purpose of the calculation of its radiation. The maximum  of the total east-west current (310 Amps) remains same for various simulation parameters due to the magnetic shielding.
\end{abstract}


\maketitle

\section{Introduction}
The short scale size (50-100 km) electromagnetic fields in the ELF range are generated below Earth's ionosphere by variety of phenomena such as lightning discharges \citep{helliwell73,uman87,inan85,berthelier08,master83,milikh95}, impulsive fields created by seismic events \citep{hayakawa06} or by ground based horizontal electric dipole antennas \citep{papadopoulos08}. These fields can interact with the E-region (90-120 km) of the Earth's ionosphere and generate currents and associated electromagnetic fields which will radiate in the presence of the conducting Earth. Infact, inspired by the physics of the quasi-static equatorial electrojet (EEJ) a novel concept of the conversion of an ELF ground based horizontal electric dipole antenna (HED) into a vertical electric dipole (VED) antenna in the E-region was proposed \citep{papadopoulos08}. At ELF frequencies, the VED radiates $10^5$ times more efficiently than HED of the same dipole current moment \citep{field89}.  In the case of quasi-static EEJ, tidal
motions drive and maintain horizontal (zonal) electric fields
of the order .5-1 mV/m perpendicular to the ambient
magnetic field over long times (several minutes to hours)
and over a 600 km strip in the E-region of the
dip equatorial ionosphere. 
This electric field drives downward Hall current.
At steady state and to
zero order, current continuity requires that a vertical polarization electric field be built to prevent the downward Hall
current from flowing. This electric field is larger than the
zonal electric field by the ratio of the Hall-to-Pedersen
conductivity, approximately a factor 30, resulting in vertical electric fields in excess of 10 mV/m and associated
predominantly with eastern electrojet currents of more than
10 A/km. As noted by Forbes [1981], current continuity
requires the presence of a vertical current, with current
closure established by field aligned currents. These currents
result in ground based quasistationary magnetic fields of
100 nT or more.

The interaction of quasistatic electric fields with the
equatorial E-region leading to the generation of EEJ  has been studied extensively both experimentally and theoretically
 \citep{kelleybook,forbes81,rastogi89,onwumechilibook,rishbeth97}. However, the interaction of ELF range electromagnetic fields with $E$-region is relatively less understood. \citet{eliasson09} studied generation and penetration of the ELF current and electromagnetic fields into the equatorial E-region. They found that the interaction of the ELF pulsed and continuous wave fields with the equatorial E-region leads to the generation of both vertical and horizontal currents which penetrate into ionospheric layers as Helicon waves. It is the objective of this paper to study the physics of the interaction of the ELF electromagnetic field  with off-equatorial E-region where the Earth's magnetic field makes finite angle $\theta$ with the horizontal. It is found that as the angle between Earth's magnetic field and horizontal is increased (going away from the equator), the  currents and fields penetrate deeper into the ionospheric layers with increased wavelength, the magnitudes of horizontal (east-west) and vertical currents decrease near the boundary and the vertical electric field decreases drastically.

The paper is organized as follows. The next section contains physics of the E-region, numerical model, boundary conditions and simulation setup. Numerical fits for the real ionospheric conductivities used in the simulations are provided in this section.  Section \ref{results} presents modelling results for various simulation parameters, viz., day-night time conditions, pulsed and continuous wave antenna field and various values of the angle between the Earth's magnetic field and the horizontal. In section \ref{summary}, we summarize our results.
\section{Interaction Model and Simulation Setup}
\begin{figure}
\begin{center}$
\begin{array}{c}
(\mathrm{a}) \\
\includegraphics[width=.6\textwidth]{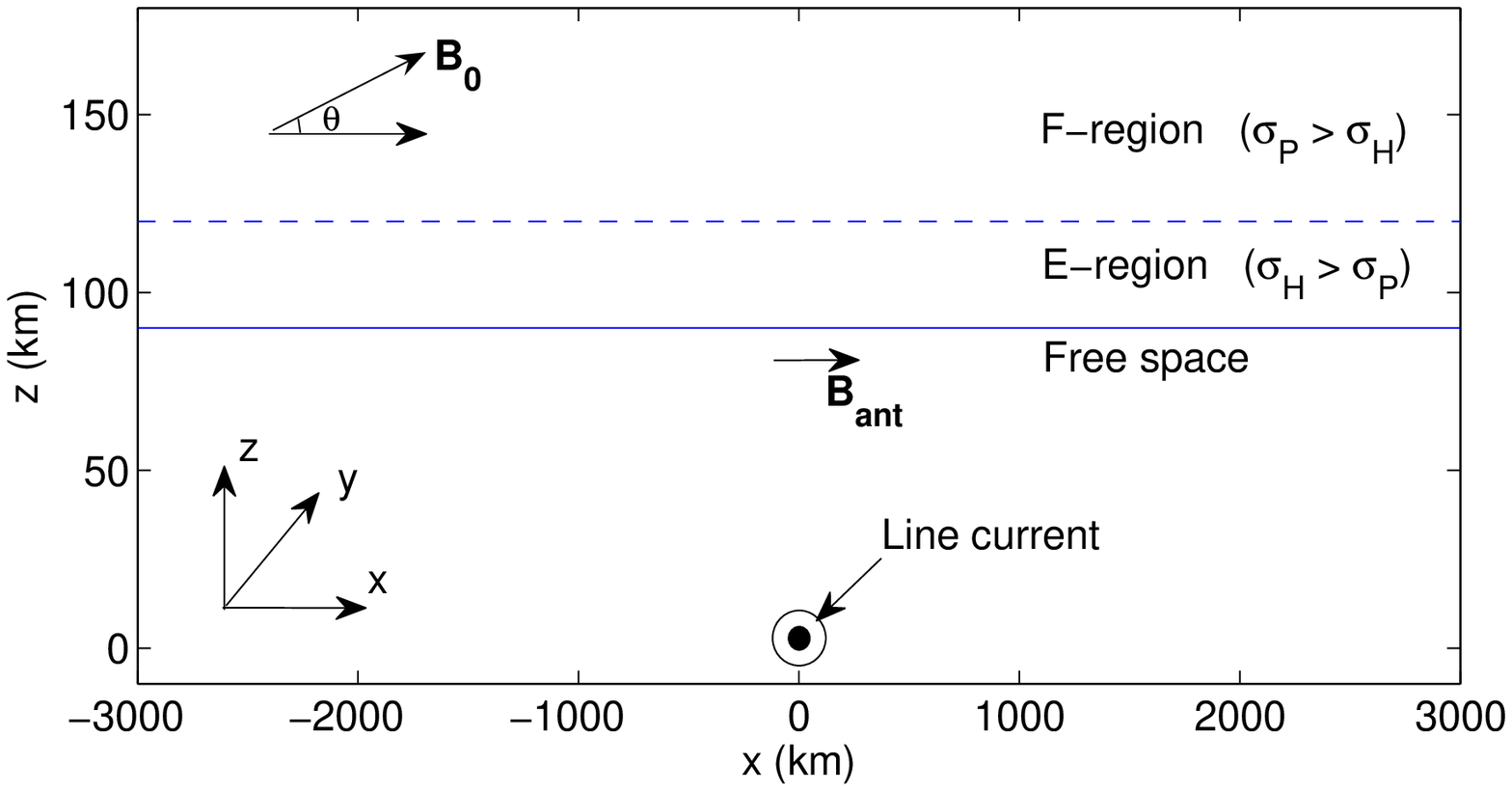}\\
(\mathrm{b})\\
\includegraphics[width=.6\textwidth]{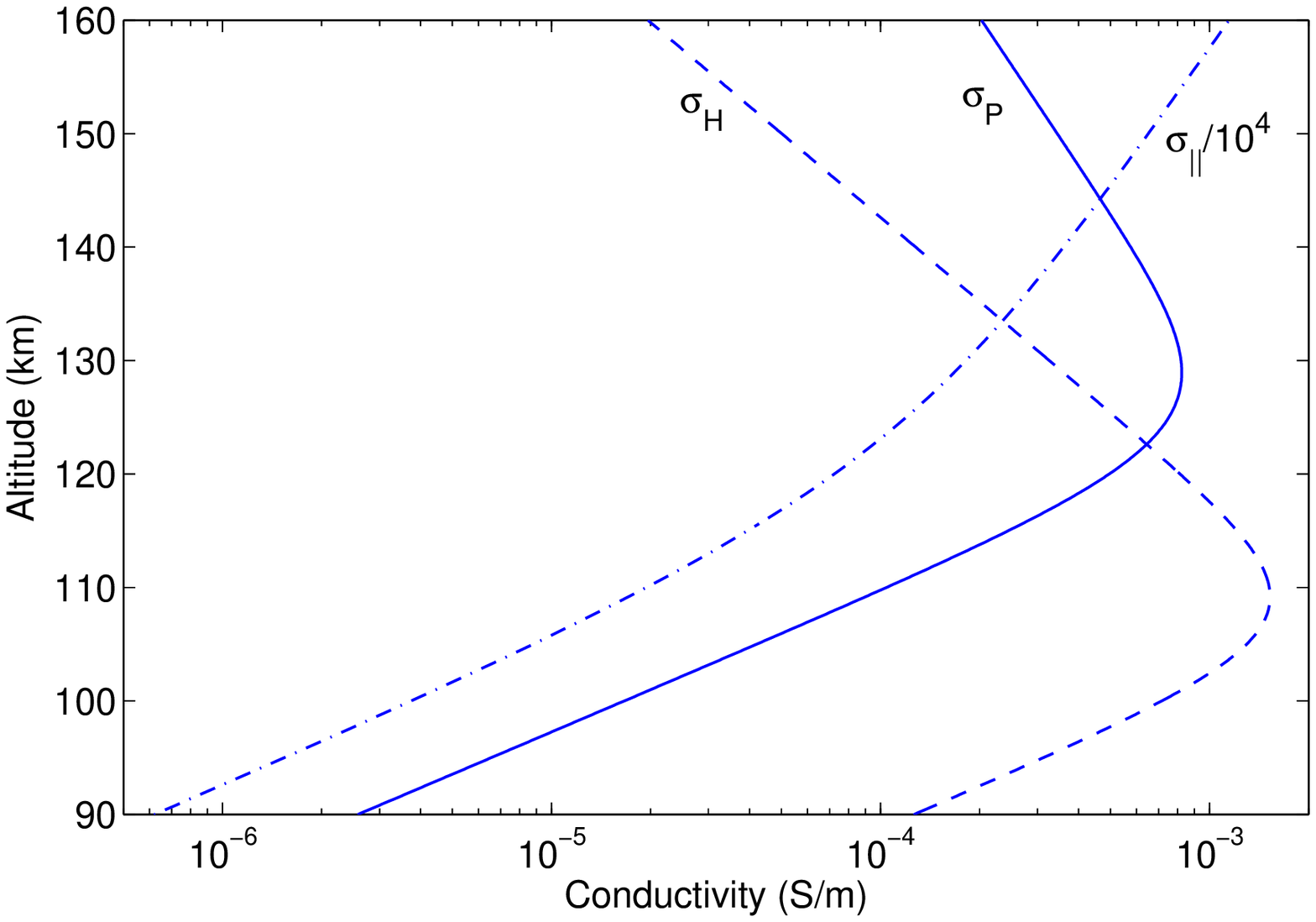}
\end{array}$
\end{center}
\caption{(a) The geometry of the simulation. The ionospheric layer is above $z=90$ km and free space is below $z=90$ km. The simulation box extends horizontally from $x=-3000$ km to $x=3000$ km and vertically  from $z=90$ km to $z=160$ km. The constant geomagnetic field $\mathbf{B}_0$ makes an angle $\theta$ with the positive $x$ direction. The ELF magnetic field $\mathbf{B}_{ant}$ which couples to the ionosphere at $z=90$ km is generated by an antenna placed at $x=z=0$. (b) Numerical fits of $\sigma_{||}$, $\sigma_{P}$ and $\sigma_H$ as given by \citet{forbes76} for day time conditions. For night time conditions, the conductivities are decreased by a factor of 5. (After \cite{eliasson09}.)}
\label{geometry_sigma}
\end{figure}

The E-region of the Earth's ionosphere (90 to 120 km above the Earth surface) consists of partially ionized plasma magnetized by the Earth's magnetic field. In this region, ions are viscously coupled to neutrals through collisions ($\nu_{in} >> \omega_{ci}$) while electrons are strongly magnetized ($\nu_{en} << \omega_{ce}$). Here $\nu_{in}$ and $\nu_{en}$  represent ion-neutral and electron-neutral collision frequencies, and  $\omega_{ci}$ and $\omega_{ce}$ represent ion and electron cyclotron frequencies respectively. The presence of the background Earth's magnetic field $\mathbf{B_0}$ drives currents  perpendicular to the magnetic field, in addition to the parallel currents. This gives rise to the tensor conductivity of the ionosphere. The current is related to the electric field by generalized Ohm's law.
\begin{equation}
 \mathbf{J}=\sigma_{||}\mathbf{E}_{||}+\sigma_P\mathbf{E}_{\perp}-\sigma_H\frac{\mathbf{E}\times\mathbf{B_0}}{B_0},
\end{equation}
where $\sigma_{||}$, $\sigma_P$ and $\sigma_H$ are parallel, Pederson and Hall conductivities respectively. The current parallel to the magnetic field $\mathbf{B_0}$ is controlled  by $\sigma_{||}$, along the perpendicular electric field $\mathbf{E}_{\perp}$ (Pederson current) by $\sigma_P$ and perpendicular to both $\mathbf{E}_{\perp}$ and $\mathbf{B_0}$ (Hall current) by $\sigma_H$. These conductivities depend on density, collision frequency and magnetic field, and are given by,
\begin{eqnarray}
 \sigma_{||}&=&\epsilon_0\left(\frac{\omega_{pe}^2}{\nu_{en}}+\frac{\omega_{pi}^2}{\nu_{in}}\right)\\
 \sigma_{P}&=&\epsilon_0\frac{\omega_{pe}^2}{\omega_{ce}}\left(\frac{\nu_{en}\omega_{ce}}{\omega_{ce}^2+\nu_{en}^2}+\frac{\nu_{in}\omega_{ci}}{\omega_{ci}^2+\nu_{in}^2}\right)\\
 \sigma_{H}&=&\epsilon_0\frac{\omega_{pe}^2}{\omega_{ce}}\left(\frac{\omega_{ce}^2}{\omega_{ce}^2+\nu_{en}^2}-\frac{\omega_{ci}^2}{\omega_{ci}^2+\nu_{in}^2}\right)
\end{eqnarray}
Here it has been assumed that the plasma consists of electrons and one ion plasma species. In the E-region, where $\omega_{ce} >> \nu_{en}$ and $\omega_{ci} << \nu_{in}$, the dominant Hall conductivity gives rise to the Helicon waves which mainly govern the plasma dynamics. While in F-region (above 120 km), where $\omega_{ce} >> \nu_{en}$ and $\omega_{ci} >> \nu_{in}$, the Hall conductivity decreases and plasma dynamics is mainly diffusive. In our numerical modeling, we have approximated the  real vertical profiles of day time conductivities \citep{forbes76} by numerical fits given by the following functions.
\begin{eqnarray}
 \sigma_{||}&=&\frac{1}{a_{1,||}\exp(-z/L_{1,||})+a_{2,||}\exp(-z/L_{2,||})}\label{sigmapar_fit}\\
 \sigma_{P}&=&\frac{1}{a_{1,P}\exp(-z/L_{1,P})+a_{2,P}\exp(z/L_{2,P})}\label{sigmaP_fit}\\
 \sigma_{H}&=&\frac{1}{a_{1,H}\exp(-z/L_{1,H})+a_{2,H}\exp(z/L_{2,H})}\label{sigmaH_fit}
\end{eqnarray}
  The  values of the parameters $a$'s and $L$'s are listed in Table \ref{table1} and the profiles for day time conditions are plotted in Fig. \ref{geometry_sigma}b. For night time conditions we assume that the conductivities are decreased by a factor of five due to the decreased electron and ion densities. It can be seen that in E-region, Hall and Pederson conductivities increase with altitude and peak around 110 km and 130 km respectively. The Hall conductivity dominates Pederson conductivity in the E-region while Pederson conductivity is dominant above 120 km.  

\begin{center}
\begin{table}[h]
\begin{tabular}{|l|l|l|l|}\hline
 $a_{1,||}=3.07\times 10^9 \Omega$m & $L_{1,||}=5.36$ km & $a_{2,||}=0.63\times 10^3 \Omega$m & $L_{2,||}=18$ km\\\hline
 $a_{1,P}=7.67\times 10^{12} \Omega$m & $L_{1,P}=5.36$ km & $a_{2,P}=0.99 \Omega$m & $L_{2,P}=18.8$ km\\\hline
 $a_{1,H}=1.53\times 10^{11} \Omega$m & $L_{1,H}=5.36$ km & $a_{2,H}=0.0157\Omega$m & $L_{2,H}=10.67$ km\\\hline
\end{tabular}
\caption{Parameter values used in the numerical fits (\ref{sigmapar_fit})-(\ref{sigmaH_fit}) of the day time conductivity profiles.}
\label{table1}
\end{table}
\end{center}

The geometry of the simulation is shown in Fig. \ref{geometry_sigma}a. The ionospheric layer is treated as a two-dimensional layer
varying in the horizontal direction $x$ and the
vertical direction $z$, and  no variation along $y$. In this model,
free space is below $z =z_0= 90$ km, while the ionospheric layer
extends vertically above $z = 90$ km. The simulation box covers the region from $z=90$ km to $z=160$ km and from $x=-3000$ km to $x=3000$ km. The ionospheric layer
and free space are magnetized by a constant external
geomagnetic field $\mathbf{B}_0$ which makes an angle $\theta$ with the positive 
$x$ direction.
 
Similar to the previous work \citep{eliasson09}, we assume a very simple model for
the antenna, that of an equivalent infinite length wire in
the $y$ direction located at $x = z = 0$, so that, 
\begin{equation}
 \mathbf{B}_{ant}(x,z,t)=B_{ant}(t)\frac{\hat{x}\bar{z}-\hat{z}\bar{x}}{(\bar{x}^2+\bar{z}^2)}.\label{bant}
\end{equation}
where $B_{ant}(t)$ is the value of the antenna magnetic field at
the bottom of the E-region ($z = z_0$) and $x=0$, and the
normalized coordinates are $\bar{x} = x/z_0, \bar{y} = y/z_0$, and $\bar{z} = z/z_0$.

\begin{figure}
 \includegraphics{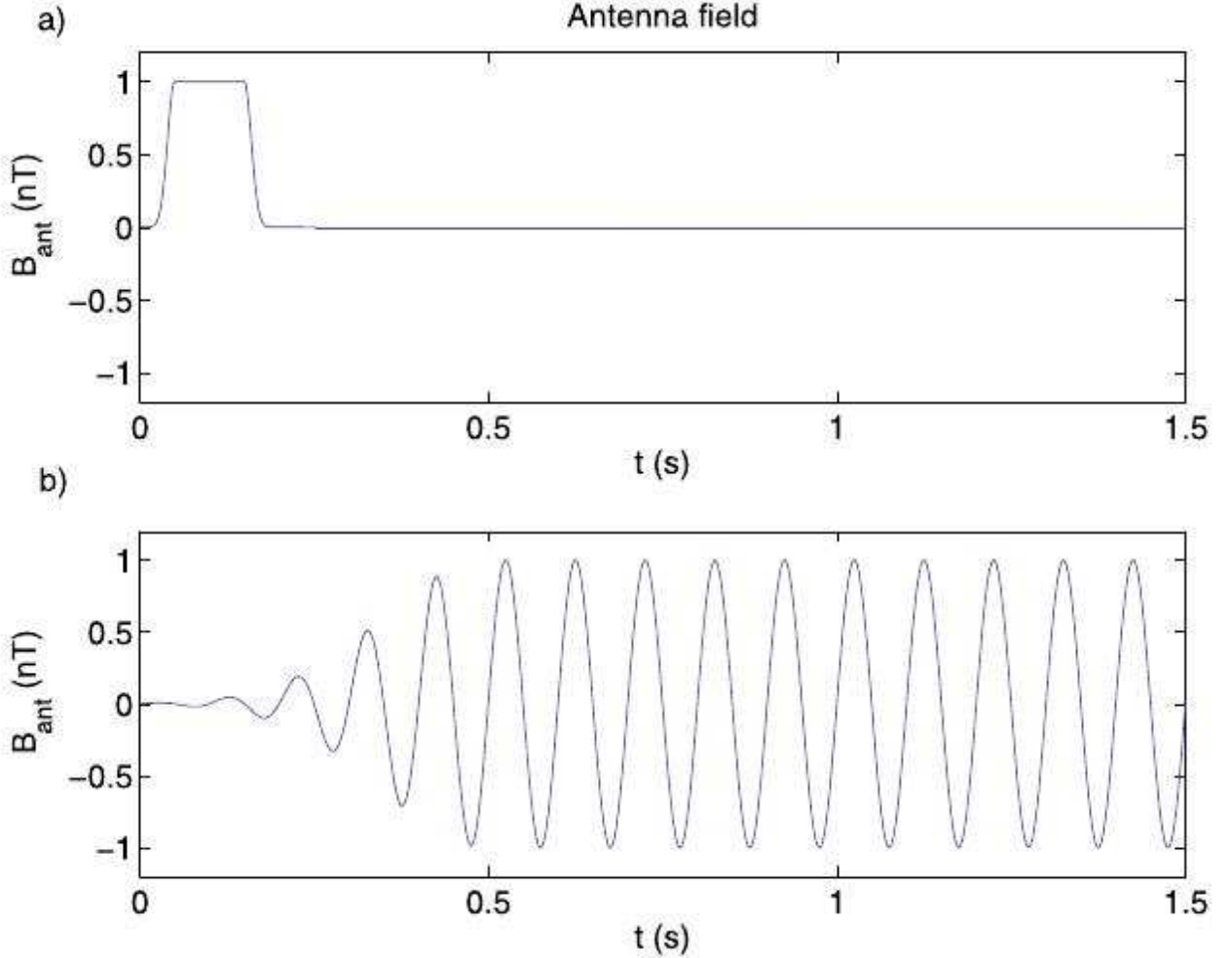}
\caption{The time dependence of the antenna magnetic field $B_{ant}(t)$ used in the simulations. (a) pulsed antenna field with pulse width 0.1 sec. (b) continuous wave antenna field with frequency 10 Hz.}
\label{empulse}
\end{figure}

We have chosen two kinds of time dependence for $B_{ant}(t)$, namely, pulsed antenna field  and continuous wave antenna field. The time dependence of
the antenna field for the two cases are shown in Fig. \ref{empulse}.
For the pulsed case we use an antenna field of the
form $B_{ant} = B_{0,ant} \exp[-(t-t_0)^2/2D_t^2]$ for $t < t_0$, $B_{ant} = B_{0,ant}$ for $t_0 \leq t < t_1$ and $B_{ant} = B_{0,ant} \exp[-(t-t_1)^2/2D_t^2]$ for $t \geq t_1$,
where the maximum amplitude $B_{0,ant} = 1$ nT is reached at
time $t_0 = 0.05$ s, and the pulse switched off smoothly at $t_1 =
0.15$ s, using the pulse rise and decay time $D_t = 0.01$ s. 
For the
continuous wave case we choose a 10 Hz antenna
field that is ramped up smoothly so that $B_{ant} = B_{0,ant}
\exp[-(t-t0)^2/2 D_t^2]\sin(20\pi t)$ for $t < t_0$ and $B_{ant} = B_{0,ant}
\sin(20\pi t)$ for $t \geq t_0$, where the pulse reaches its maximum
amplitude $B_{0,ant} = 1$ nT at $t_0 = 0.5$ s, and the rise time is $D_t =
0.15$ s.

It is convenient for numerical purpose to express the generalized Ohm's law in terms of the impedance tensor. 
\begin{equation}
 \mathbf{E}=\bar{\bar{\rho}}\mathbf{J}
\end{equation}
where $\bar{\bar{\rho}}$ is the impedance tensor  which is obtained by inverting the conductivity tensor. For the background magnetic field making an angle $\theta$ with positive $x$-direction, it is given by,\\
\[
 \bar{\bar{\rho}}=\left[\begin{array}{ccc}
 \rho_{||}\cos^2\theta+\rho_P\sin^2\theta&\rho_H\sin\theta&(\rho_{||}-\rho_P)\sin\theta\cos\theta\\
 -\rho_H\sin\theta&\rho_P&\rho_H\cos\theta\\
(\rho_{||}-\rho_P)\sin\theta\cos\theta&-\rho_H\cos\theta& \rho_{||}\sin^2\theta+\rho_P\cos^2\theta
                  \end{array}\right]
=\left[\begin{array}{ccc}
\rho_{11}&\rho_{12}&\rho_{13}\\
-\rho_{12}&\rho_{22}&\rho_{23}\\
\rho_{13}&-\rho_{23}&\rho_{33}
\end{array}
\right]\]\\
where $\rho_{||}=1/\sigma_{||}$, $\rho_{P}=\sigma_P/(\sigma_P^2+\sigma_H^2)$ and $\rho_{H}=\sigma_H/(\sigma_P^2+\sigma_H^2)$. The other equations governing the dynamics in the E-layer are Ampere's law and Farady's law. 
\begin{eqnarray}
 \nabla\times\mathbf{B}&=&\mu_0\mathbf{J}\\
\frac{\partial \mathbf{B}}{\partial t}&=&-\nabla\times\mathbf{E}
\end{eqnarray}
Combining Ampere's and Faraday's law with Ohm's law, we obtain an evolution equation for  $\mathbf{B}$.
\begin{equation}
 \frac{\partial \mathbf{B}}{\partial t}=-\frac{1}{\mu_0}\nabla\times[\bar{\bar{\rho}}.\nabla\times\mathbf{B}]
\label{evol_eq}
\end{equation}
 Equation (\ref{evol_eq}) describes the Helicon wave dynamics when impedance tensor is nondiagonal and diffusive phenomena when impedance tensor is diagonal. Separating equation (\ref{evol_eq}) into components and using the condition $\nabla.\mathbf{B}=0$, we get three coupled equations in $B_x$, $B_y$ and $B_z$.
\begin{eqnarray}
 \frac{\partial B_y}{\partial t}&=&-\frac{1}{\mu_0} 
\frac{\partial}{\partial z}\left\{ -\rho_{11}\frac{\partial B_y}{\partial z}+\rho_{12}\left(\frac{\partial B_x}{\partial z}-\frac{\partial B_z}{\partial x}\right)+\rho_{13}\frac{\partial B_y}{\partial x}\right\}\nonumber\\
&&+\frac{1}{\mu_0}\frac{\partial}{\partial x}\left\{ -\rho_{13}\frac{\partial B_y}{\partial z}-\rho_{23}\left(\frac{\partial B_x}{\partial z}-\frac{\partial B_z}{\partial x}\right)+\rho_{33}\frac{\partial B_y}{\partial x}\right\}\label{evol_by}\\
\frac{\partial B_z}{\partial t}&=&-\frac{1}{\mu_0}\frac{\partial}{\partial x}\left\{ \rho_{12}\frac{\partial B_y}{\partial z}+\rho_{22}\left(\frac{\partial B_x}{\partial z}-\frac{\partial B_z}{\partial x}\right)+\rho_{23}\frac{\partial B_y}{\partial x}\right\}\label{evol_bz}\\
\frac{\partial B_x}{\partial x}&=&-\frac{\partial B_z}{\partial z}\label{divb}
\end{eqnarray}
 Here $\nabla.\mathbf{B}=0$ is used in place  of the $x$-component  of the evolution equation (\ref{evol_eq}) to calculate $B_x$. For $\theta=0^o$, equations (\ref{evol_by})-(\ref{divb}) and impedance tensor $\bar{\bar{\rho}}$ reduce to the forms which were studied earlier by \citet{eliasson09}.

As boundary condition at the plasma-free space boundary $z=z_0$, we use the continuity of the magnetic field $B_z$ and its z-derivative \citep{eliasson09}.
This also gives continuity of the parallel (to the boundary surface in the $x$-$y$ plane) electric field. A detailed discussion about the boundary conditions is given by \cite{eliasson09}.
%

\section{Results\label{results}}
We have conducted a series of numerical studies of
the system (\ref{evol_by})-(\ref{divb}) with various parameters of interest, viz., day and night time conductivities, pulsed and continuous wave antenna field and different  values of $\theta$. 
In what follows we first present results for a fixed value of $\theta=5^o$ and vary other parameters. Then we change values of  $\theta$ for all other parameters in order to see the $\theta$-dependence of the  results. Finally, a distribution will be fitted to $z$-integrated $J_y$ profile, which is useful for the calculation of the radiation from $J_y$. 
\begin{figure}  \includegraphics[width=\textwidth,height=.5\textheight]{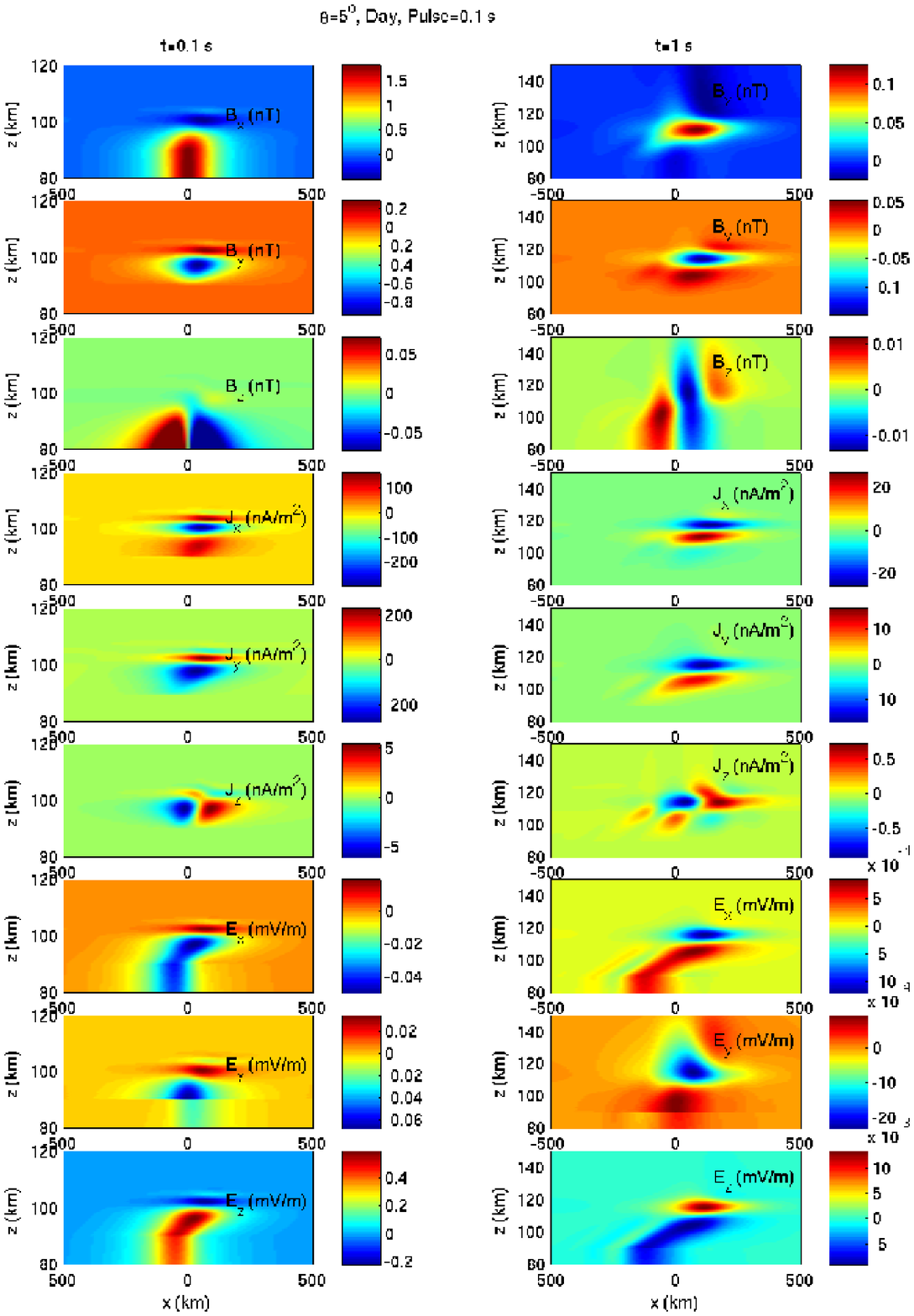}
\caption{The magnetic field components $B_x$, $B_y$ and $B_z$ (nT), the current density components $j_x$, $j_y$ and $j_z$ (nA/m$^2$) and the electric field components $E_x$, $E_y$ and $E_z$ (mV/m) for day time conditions, pulsed antenna field and $\theta=5^o$ at $t=0.1$ sec. (left column) and $t=1$ sec. (right column)\label{var_theta5_pulse_day}}.
 \end{figure}
\begin{figure}  \includegraphics[width=\textwidth,height=.5\textheight]{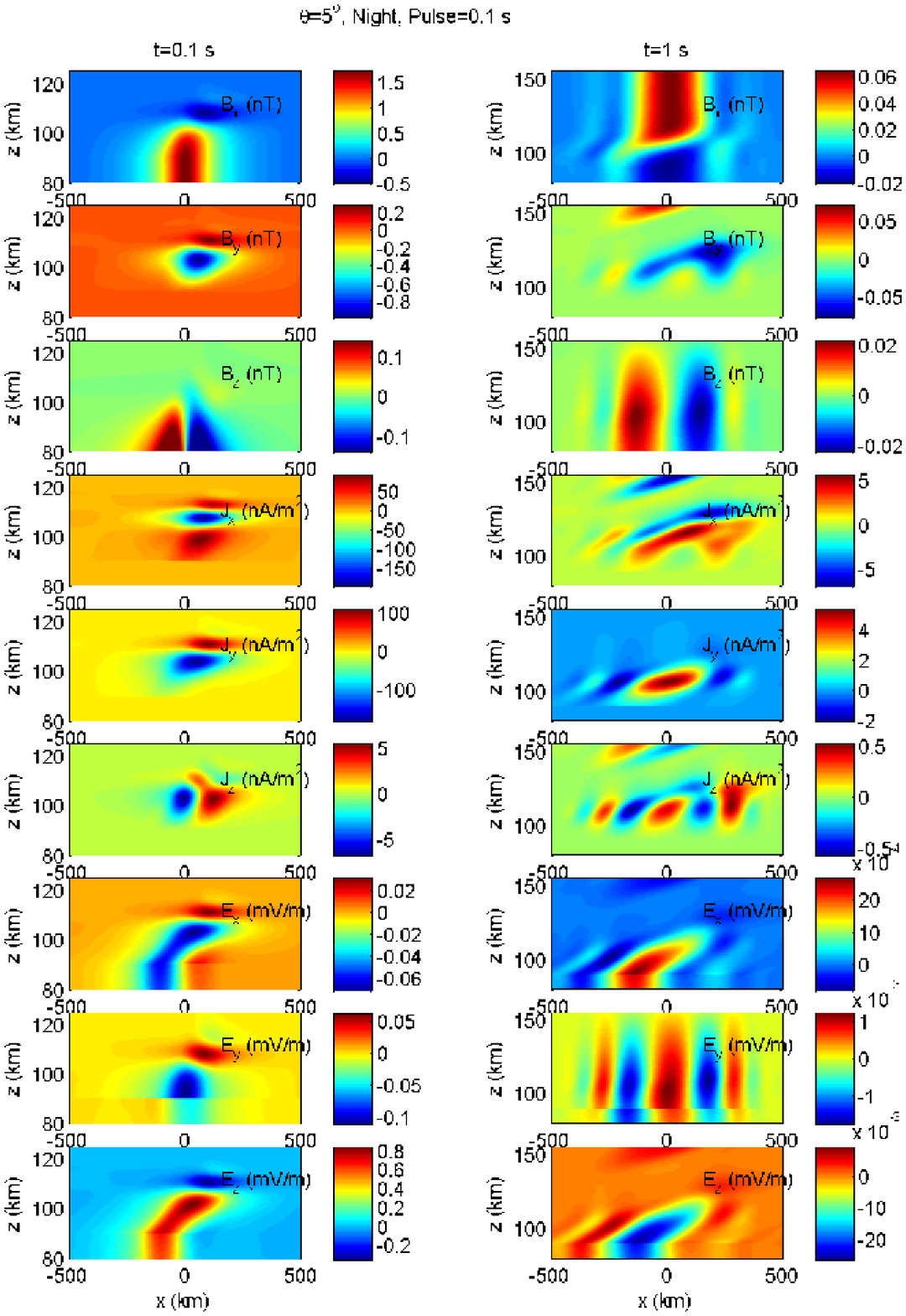}
\caption{The magnetic field components $B_x$, $B_y$ and $B_z$ (nT), the current density components $j_x$, $j_y$ and $j_z$ (nA/m$^2$) and the electric field components $E_x$, $E_y$ and $E_z$ (mV/m) for night time conditions, pulsed antenna field and $\theta=5^o$ at $t=0.1$ sec. (left column) and $t=1$ sec. (right column)\label{var_theta5_pulse_night}}.
 \end{figure}

Figs. \ref{var_theta5_pulse_day} and \ref{var_theta5_pulse_night} show spatial profiles of magnetic field, currents and electric fields  for pulsed antenna field for the case $\theta=5^o$ for day and night time conductivities, respectively.  The left column shows results at $t=0.1$ s when the antenna field is at its maximum (1 nT), and right column in the relaxation phase at $t=1$ s when the antenna field is zero. 
It can be seen at $t=0.1$ s that interaction of the antenna field with the ionospheric plasma at the lower boundary at $z=90$ km generates currents and associated electromagnetic fields structures. The oblique wave penetration with decreasing wavelength of these structures into the deeper ionospheric layers up to $z \sim 110$ km can also be seen. The $x$-scale $\sim 100$ km of the structures is larger than the $z$-scale $\sim 10$ km. The amplitude of the vertical electric field $E_z \sim .6$ mV/m is one order of magnitude larger than those of the horizontal components $E_x$ and $E_y$. The vertical current $J_z$ has upward and downward  flow structure and its amplitude is 1-2 order of magnitude smaller than those of the horizontal currents $J_x$ and $J_y$. The disparity between the magnitudes of $J_x$ and $J_z$, and $x$- and $z$-scales is consistent with the current continuity condition $\partial J_x/\partial x+\partial J_z/\partial z=0$. Another interesting feature to note here is that at $x=0$ and just below $z=90$ km the horizontal magnetic field component $B_x\sim 2$ nT which is almost double of the antenna magnetic field (1 nT), while vertical magnetic field component $B_z$ is much smaller than the antenna field. This happens due to the shielding of plasma interior from the incident magnetic field.
At the lower boundary, the antenna electric field $E_y$ drives Pederson current $J_{py}=\sigma_PE_y$ along $E_y$ and Hall current $\mathbf{J}_{H\perp}$
in the $x$-$z$ plane perpendicular to the background oblique magnetic field $\mathbf{B_0}=B_0(\hat{x}\cos\theta+\hat{z}\sin\theta)$. A perpendicular electric field $\mathbf{E}_{\perp}$
develops to limit the magnitude of the $\mathbf{J}_{H\perp}$ to the value required by the current continuity condition. 
This $\mathbf{E}_{\perp}$ drives Hall current $J_{hy}=\sigma_HE_{\perp}$ in the negative $y$-direction. 
The total current $J_y=J_{hy}+J_{py}$ generates the magnetic field component $B_x$ which cancels the antenna field inside the plasma while at the ionospheric boundary at $z=90$ km and $x=0$, it adds to the antenna field to almost double its value. The $z$-component of the antenna magnetic field is partially cancelled by the magnetic field produced by the  induced current $J_y$.
 
Due to the current continuity, the perpendicular current  is shunted parallel  to $\mathbf{B_0}$.  The $z$-component of this parallel current and of the Hall current $\mathbf{J}_{H\perp}$ give rise to the upward-downward flow structure of $J_z$. The current components in terms of electric field components can be written as,
\begin{eqnarray}
 J_x&=&(\sigma_{||}\cos^2\theta+\sigma_P\sin^2\theta)E_x-\sigma_H\sin\theta E_y+(\sigma_{||}-\sigma_P)\sin\theta\cos\theta E_z\\
J_y&=&\sigma_H \sin\theta E_x+\sigma_P E_y -\sigma_H \cos\theta E_z\\ J_z&=&(\sigma_{||}-\sigma_P)\sin\theta\cos\theta E_x+\sigma_H\cos\theta E_y+(\sigma_{||}\sin^2\theta+\sigma_P\cos^2\theta)E_z
\end{eqnarray}
In the region 90 km $<z<$ 110 km,  $\sigma_{||}/\sigma_H \sim 50-150 >> 1$ and $ \sigma_{||}/\sigma_P \sim 2400-1900 >> 1$. Therefore  for small values of $\theta$ (for example for $\theta=5^o,\cos\theta \sim 1$ and $\sin\theta \sim 0.08$) $J_x$ is mainly dominated by parallel current as $E_x \sim E_y \sim 0.1 E_z$, while $J_z$ has contributions both from parallel and Hall currents in $x$-$z$ plane. For large values of $\theta$ close to $90^o$, $J_x$ and $J_z$ exchange their roles. The out-of-plane current $J_y$ is mainly Hall dominated as $\sigma_H >> \sigma_P$ for $z<120$ km  and $E_z >> E_y$.

The phase speed of the wave penetration can be estimated from whistler dispersion relation $v_{\phi}=\omega/k=k_{||}\rho_H/\mu_0$. This gives $v_{\phi}= 100$ km/s for an average value of $\rho_H=4000$ S/m in the region $90$ km $<z<$ $110$ km and a half wavelength of 100 km giving $k_{||}=\pi/100=0.314$ km$^{-1}$. This gives penetration distance $\sim 10$ km by $t=0.1$ sec., as is seen in Fig. \ref{var_theta5_pulse_day}. From whistler dispersion relation, $\lambda \propto \rho_H \sim 1/\sigma_H$ for  given values of $\omega$ and $k_{||}$. Since $\sigma_H$ increases from $z=90$ km to $z=110$ km, the wavelength decreases, which is consistent with the simulations. 

 In the relaxation phase ($t=1$ sec.) when antenna field has decayed to zero, the sign of the structures have reversed and the magnitudes have decreased. The sign is reversed because in the decay phase of the incident pulse the antenna electric field changes its sign, which generates currents and electromagnetic fields in the opposite direction. The waves penetrates up to $z \approx 120$ km, beyond which Pederson conductivity dominates Hall conductivity and penetration becomes mainly diffusive. 

For night time conditions (Fig. \ref{var_theta5_pulse_night}), the currents and electromagnetic fields penetrate deeper up to $z \sim$ 120 km by $t=0.1$ sec. and vertical scale length of the structures increases. This is due to the decreased (by a factor of 5) conductivities which increase the wavelength and phase speed of the waves as $\lambda,v_{\phi}\propto 1/\sigma_H$. The amplitudes of $J_x$ and $J_y$ decrease but not by factor of 5 as electric field amplitudes increase. In the relaxation phase, the structures have penetrated into the diffusive F-layer (above 120 km) up to the upper boundary of the simulation domain. The helicon waves with a typical wavelength of $\sim 300$ km propagate laterally in both directions. These waves are concentrated around the  Hall conductivity maximum $z=110$ km, which seems to guide the helicon waves.  

\begin{figure}
\includegraphics[width=\textwidth,height=.5\textheight]{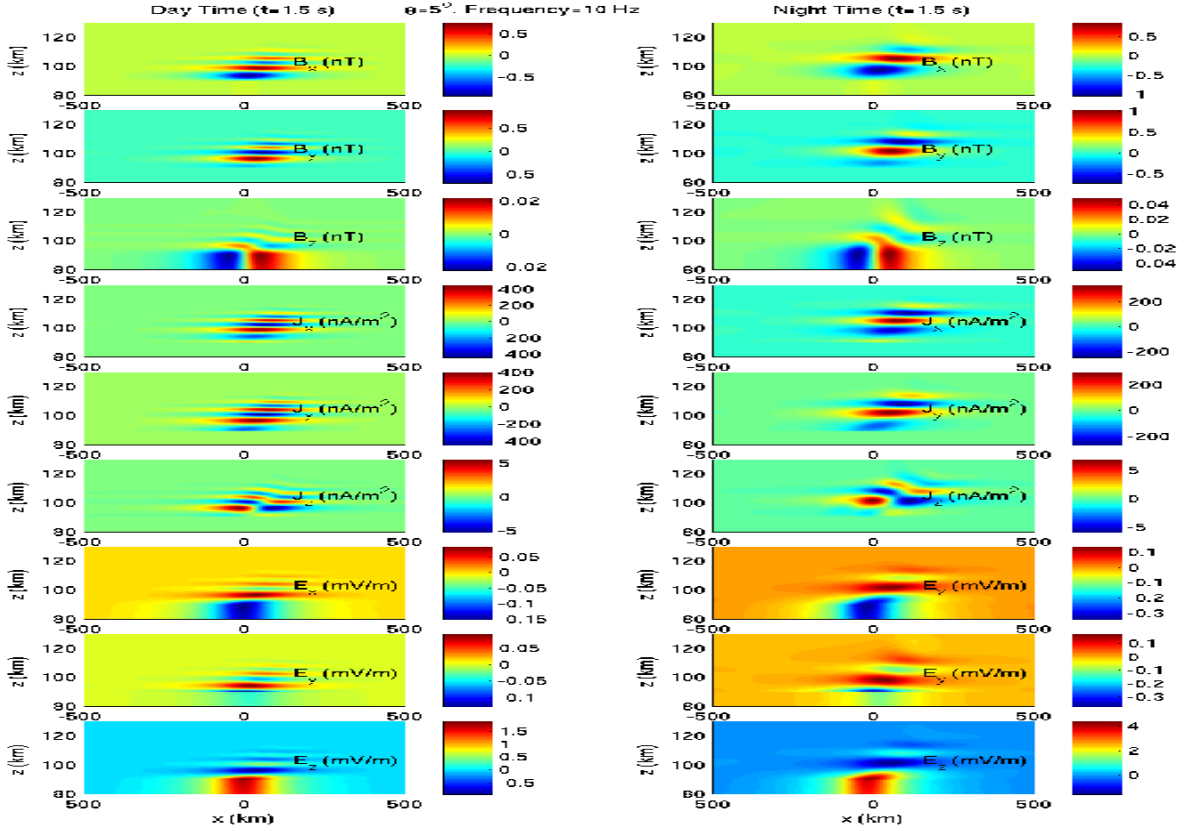}
\caption{The magnetic field components $B_x$, $B_y$ and $B_z$ (nT), the current density components $j_x$, $j_y$ and $j_z$ (nA/m$^2$) and the electric field components $E_x$, $E_y$ and $E_z$ (mV/m) for continuous wave  antenna field and $\theta=5^o$ in the steady oscillatory state ($t=1.5$ sec.). The left and right columns are for day and night time conditions respectively.  \label{var_theta5_freq_daynight}}.
 \end{figure}
Fig. \ref{var_theta5_freq_daynight} shows results for continuous wave antenna field of frequency 10 Hz and $\theta=5^o$ in the steady oscillatory state at $t=1.5$ sec. This figure shows features similar to the case of the pulsed antenna field, viz., vertical wave penetration with decreasing wavelength and deeper penetration with larger wavelength for night time conductivities as compared to day time conductivities. However,  the vertical scale length is relatively smaller than that in the case of pulsed antenna field. The reason is that in the case of pulsed antenna field the Fourier spectrum of the antenna field consists  of frequencies smaller than the frequency of the continuous wave antenna field. It can be seen from the wave dispersion relation that $\lambda \propto 1/\omega$ which gives larger wavelengths corresponding to the lower frequencies  in the pulsed antenna field case. 
\begin{figure}
\includegraphics[width=\textwidth,height=.5\textheight]{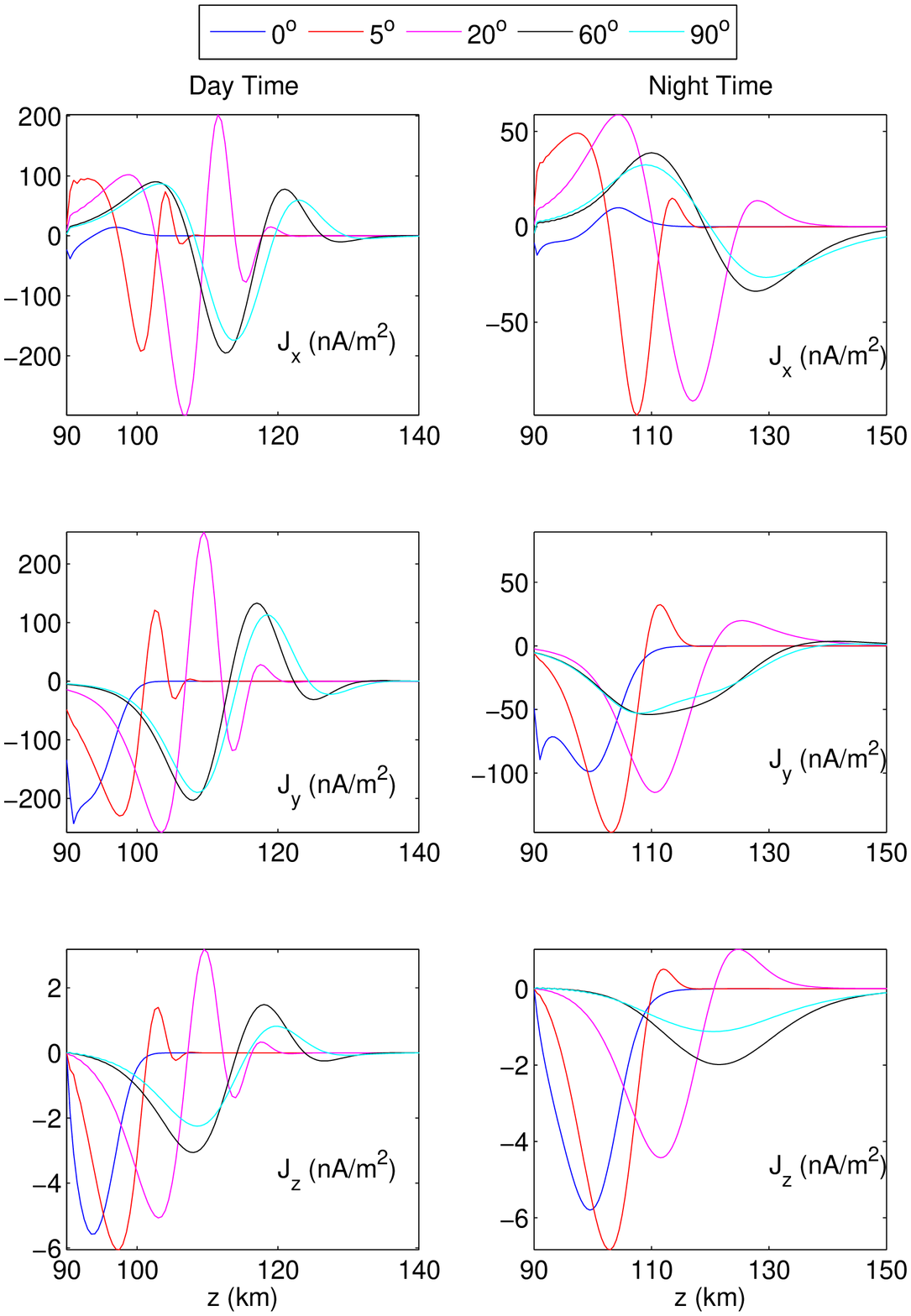}
\caption{Vertical profiles of the current density components along the line $x=0$ at $t=0.1$ sec. for pulsed  antenna field  and various values of $\theta$.  The left and right columns are for the day and night time conditions respectively.  \label{J_pulse}}.
 \end{figure}
\begin{figure}
\includegraphics[width=\textwidth,height=.5\textheight]{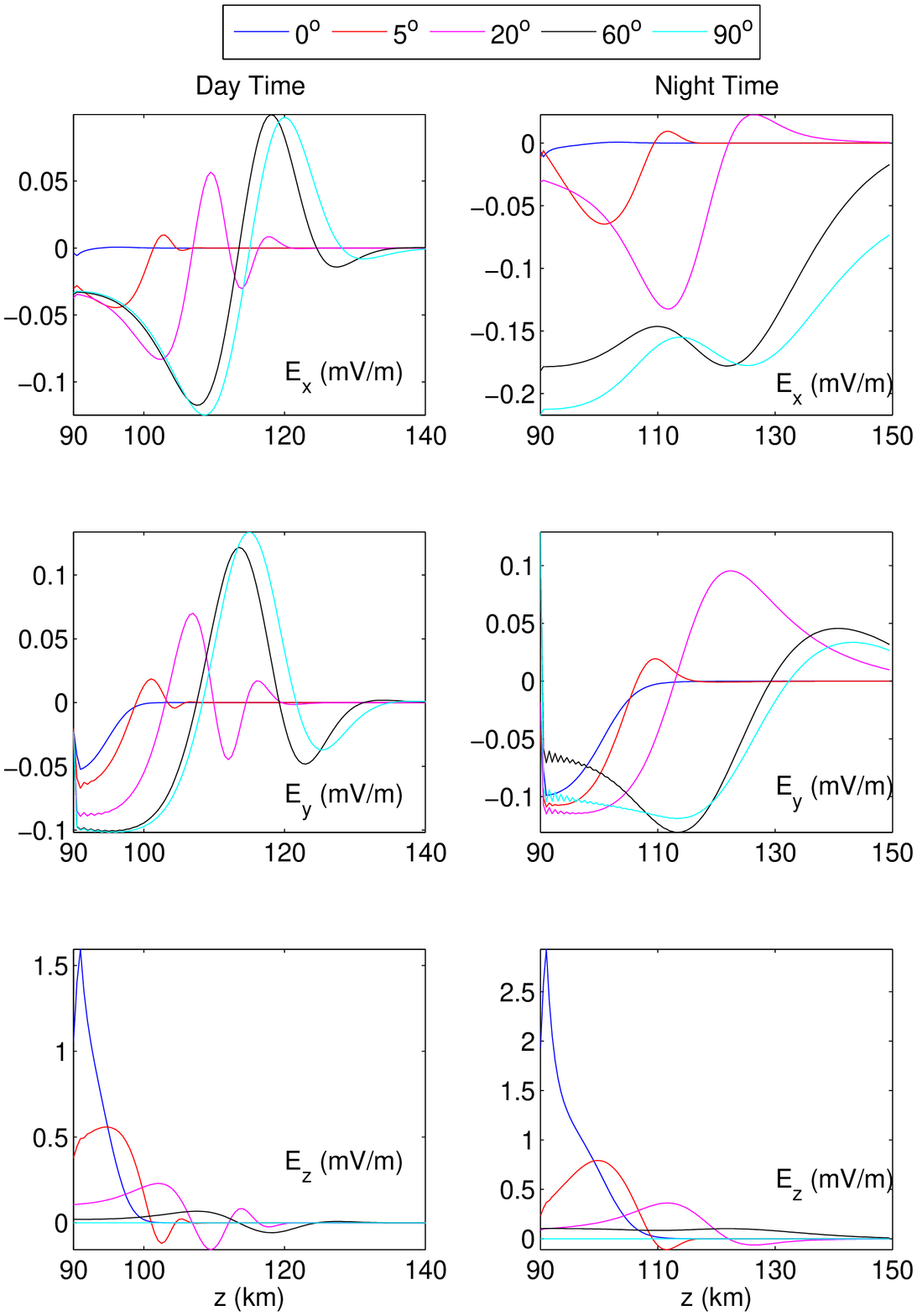}
\caption{Vertical profiles of the electric field components along the line $x=0$ at $t=0.1$ sec. for pulsed  antenna field  and various values of $\theta$.  The left and right columns are for the day and night time conditions respectively.  \label{E_pulse}}.
 \end{figure}
\begin{figure}
\includegraphics[width=\textwidth,height=.5\textheight]{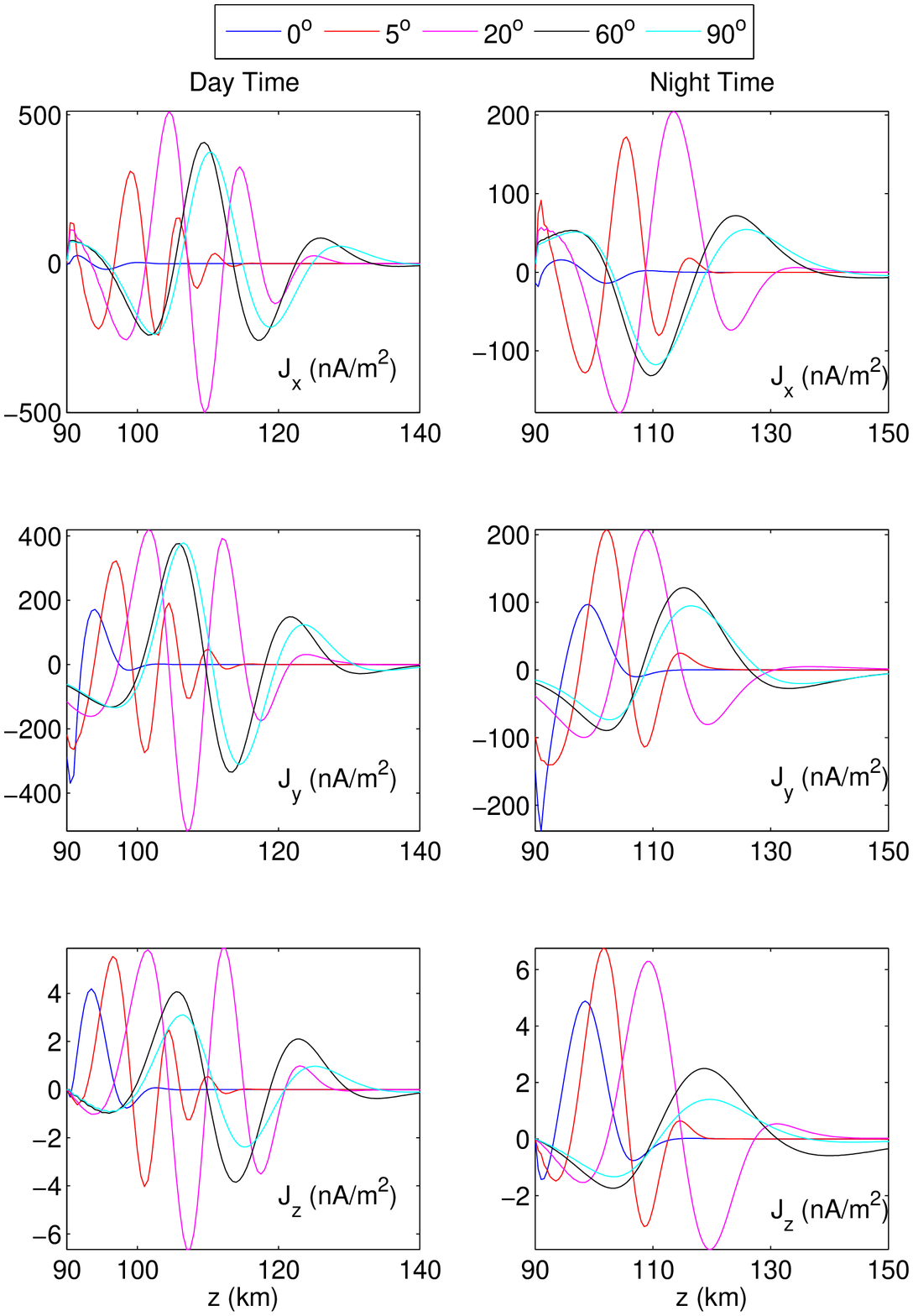}
\caption{Vertical profiles of the current density components along the line $x=0$ at $t=1.5$ sec. for continuous wave  antenna field  and various values of $\theta$.  The left and right columns are for the day and night time conditions respectively.  \label{J_freq}}.
 \end{figure}
\begin{figure}
\includegraphics[width=\textwidth,height=.5\textheight]{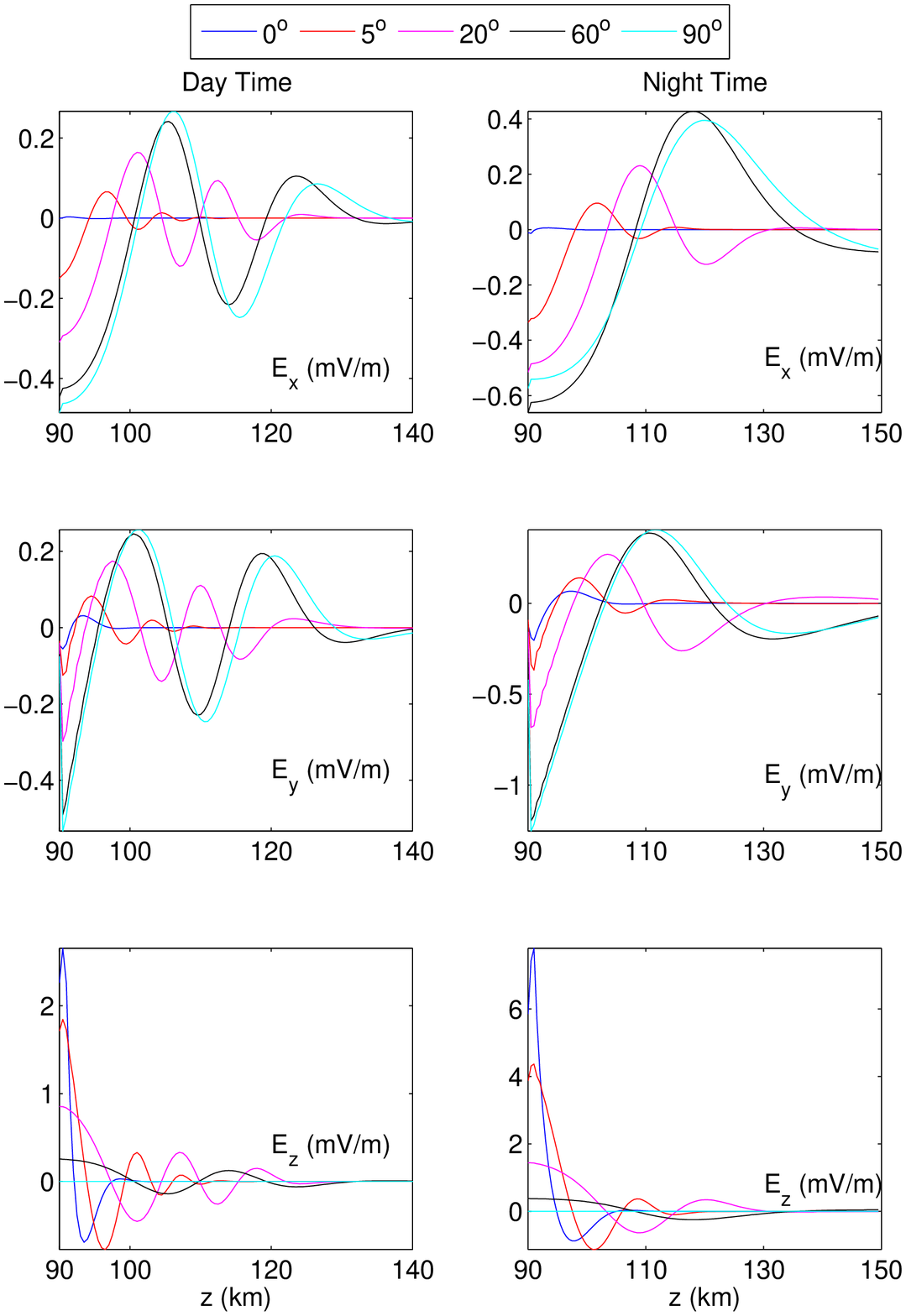}
\caption{Vertical profiles of the electric field components along the line $x=0$ at $t=1.5$ sec. for continuous wave  antenna field  and various values of $\theta$.  The left and right columns are for the day and night time conditions respectively.  \label{E_freq}}.
 \end{figure}
\subsection{$\theta$-dependence}
Now we present results by varying $\theta$ in order to see how the features observed for a single value of $\theta=5^o$ change. The vertical profiles (along line $x=0$) of the current density and electric field components are shown for various values of $\theta$ in Figs. \ref{J_pulse} and \ref{E_pulse} (pulsed antenna field), and in Figs. \ref{J_freq} and \ref{E_freq} (continuous wave antenna field). The first thing that can immediately be noted from these figures is that the vertical scale length of the structures increases with $\theta$; the structures become broader for larger values of $\theta$. Consequently, the peak of the structures shift away from the boundary ($z=90$ km) and deeper pentration takes place with increasing values of $\theta$. The magnitudes of $J_y$ and $J_z$ decrease with $\theta$ near the lower boundary. The vertical scale increases rapidly for up to $\theta=20^o$, but it is similar for $\theta=60^o$ and $90^o$ indicating asymptotic scale for large value of $\theta$. For $\theta=60^o$ and $90^o$, the profiles of $J_x$ and $J_y$  almost follow each other  while profiles of $J_z$ have similar scale but differ in magnitude. The profiles of $E_x$ and $E_y$ also almost follow each other for $\theta=60^o$ and $90^o$. 
The amplitudes of the spatial oscillations of $E_x$ and $E_y$ increase with $\theta$, while that of $E_z$ decreases very rapidly. For example, near the lower boundary $z=90$ km, for pulsed antenna field and day time conditions $E_y$ increases from $\sim -.05$ mV/m (for $\theta=0^o$) to $\sim -0.1$ mV/m (for $\theta=90^o$) and for continuous wave antenna field and day time conditions, it increases from $\sim -.05$ mV/m (for $\theta=0^o$) to $\sim -.5$ mV/m (for $\theta=90^o$). On the other hand for pulsed antenna field and day time conditions $E_z$ decreases drastically from $\sim 1.5$ mV/m for $\theta=0^o$ to very small values $\sim 10^{-4}$ mV/m for $\theta=90^o$ and for continuous wave antenna field and day time conditions from $\sim 3$ mV/m for $\theta=0^o$ to  $\sim 10^{-4}$ mV/m for $\theta=90^o$.

\begin{figure}
\includegraphics[width=\textwidth,height=.5\textheight]{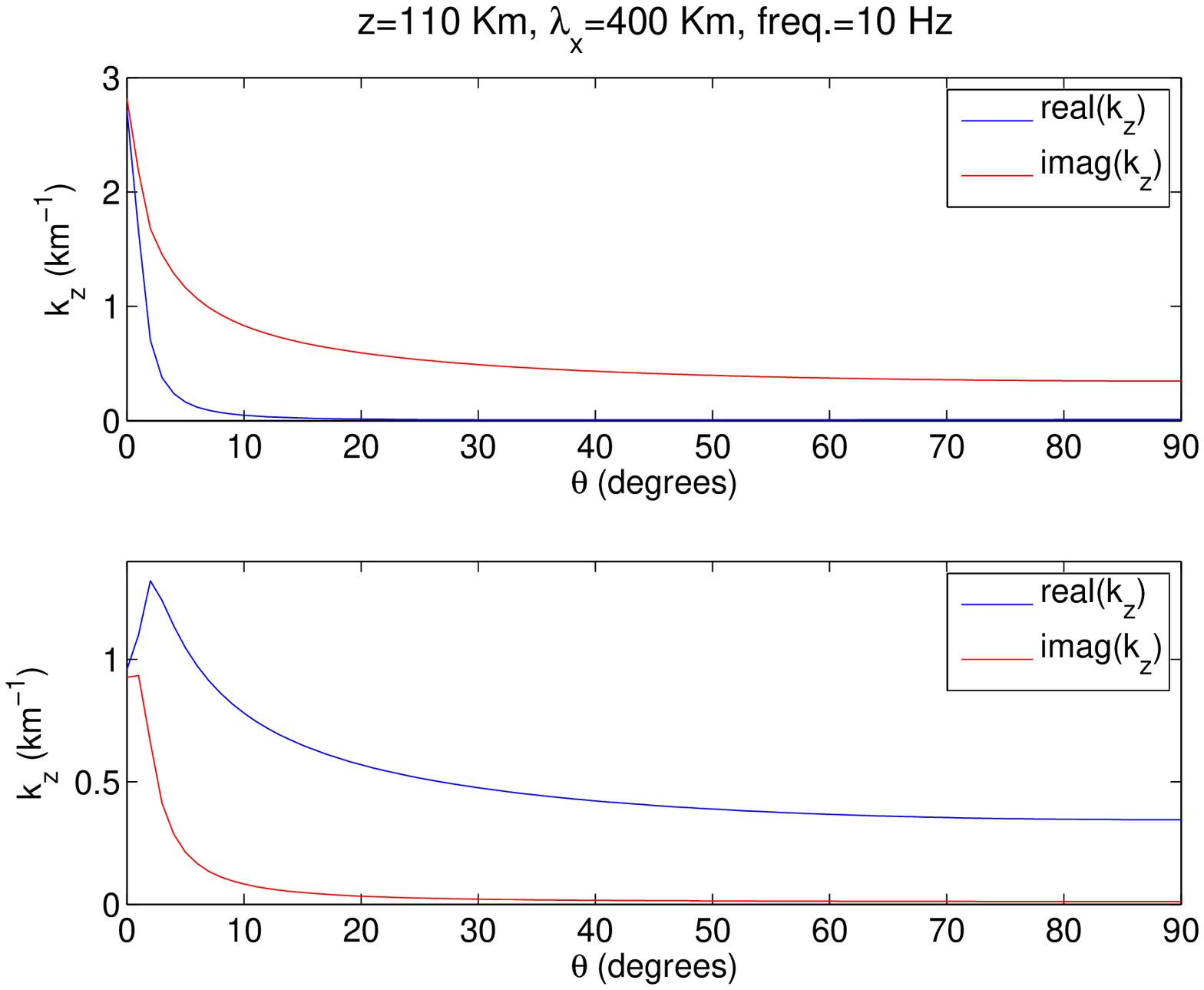}
\caption{$\theta$-dependence of the real and imaginary parts of the vertical wave number $k_z$ obtained from the solution of the dispersion relation (\ref{dispersion}) for given values of $\omega=20\pi$ s$^{-1}$, $k_{x}=\pi/200$ km$^{-1}$ and values of $\rho_P,\rho_H$ and $\rho_0$ at $z=110$ km. \label{dispfig}}.
 \end{figure}

 In order to understand the $\theta$-dependence, we have derived a local dispersion relation from Eqs. (\ref{evol_by})-(\ref{divb}) assuming uniform conductivities. Many of the observed features can be explained  qualitatively from such a  dispersion relation displayed below.
\begin{eqnarray}
 \mu_0^2\omega^2+i\omega\mu_0[k_{\perp}^2(\rho_0+\rho_P)+2k_{||}^2\rho_P]&=&(k_{\perp}^2+k_{||}^2)[k_{\perp}^2\rho_P\rho_0+k_{||}^2(\rho_P^2+\rho_H^2)],\label{dispersion}
\end{eqnarray}
where $\omega$ is frequency, $k_{\perp}=k_z\cos\theta-k_x\sin\theta$ and   $k_{||}=k_x\cos\theta+k_z\sin\theta$ ($k_x$ and $k_z$ are wave numbers along $x$ and $z$ directions respectively). In the collisionless limit ($\rho_0\rightarrow 0, \rho_P \rightarrow 0$), this dispersion relation reduces to the whistler dispersion relation $\mu_0\omega=kk_{||}\rho_H$. We have solved the dispersion relation (\ref{dispersion}) numerically for $k_z$ for the given values of $\omega=20\pi$ sec.$^{-1}$,  $k_x=\pi/200$ km$^{-1}$ and values of $\rho_0,\rho_P$ and $\rho_H$ at $z=110$ km where the Hall conductivity is maximum. The dispersion relation (\ref{dispersion}) is a fourth order polynomial in $k_z$ and hence its numerical solution gives four roots for $k_z$. The two of the four roots are the complex conjugate of the other two. We discard the roots with negative imaginary parts as they give the solution exponentially growing in $z$. The real and imaginary parts of the other two roots as functions of $\theta$ are plotted in top (root 1) and bottom (root 2) panels of Fig. \ref{dispfig}. As $\theta$ increases, the real and imaginary parts of both the roots approach asymptotic values for large values of $\theta$. For root 1, real part of $k_z$ drops down to very small values  while imaginary part $\sim 0.5$ km$^{-1}$ for large values of $\theta$. This root has very long wavelength but decays in a small distance $\sim 2$ km. 
For root 2, imaginary part of $k_z$ attains very small values while real part $\sim 0.4$ km$^{-1}$ for large values of $\theta$, giving a large decaying distance and a wavelength $\sim 15$ km. This root will be visible in the simulations. The real part for root 2 decreases rapidly up to $\theta\sim 20^o$ and then slowly up to $\theta=90^o$. This is consistent with the simulations in which vertical scale length increases significantly up to $\theta=20^o$ but there is not much difference in the vertical scales for large values of $\theta$, e.g., scale lengths for $\theta=60^o$ and $90^o$ in Figs. \ref{J_pulse}-\ref{E_freq}.    

\begin{figure}
\includegraphics[width=\textwidth,height=.5\textheight]{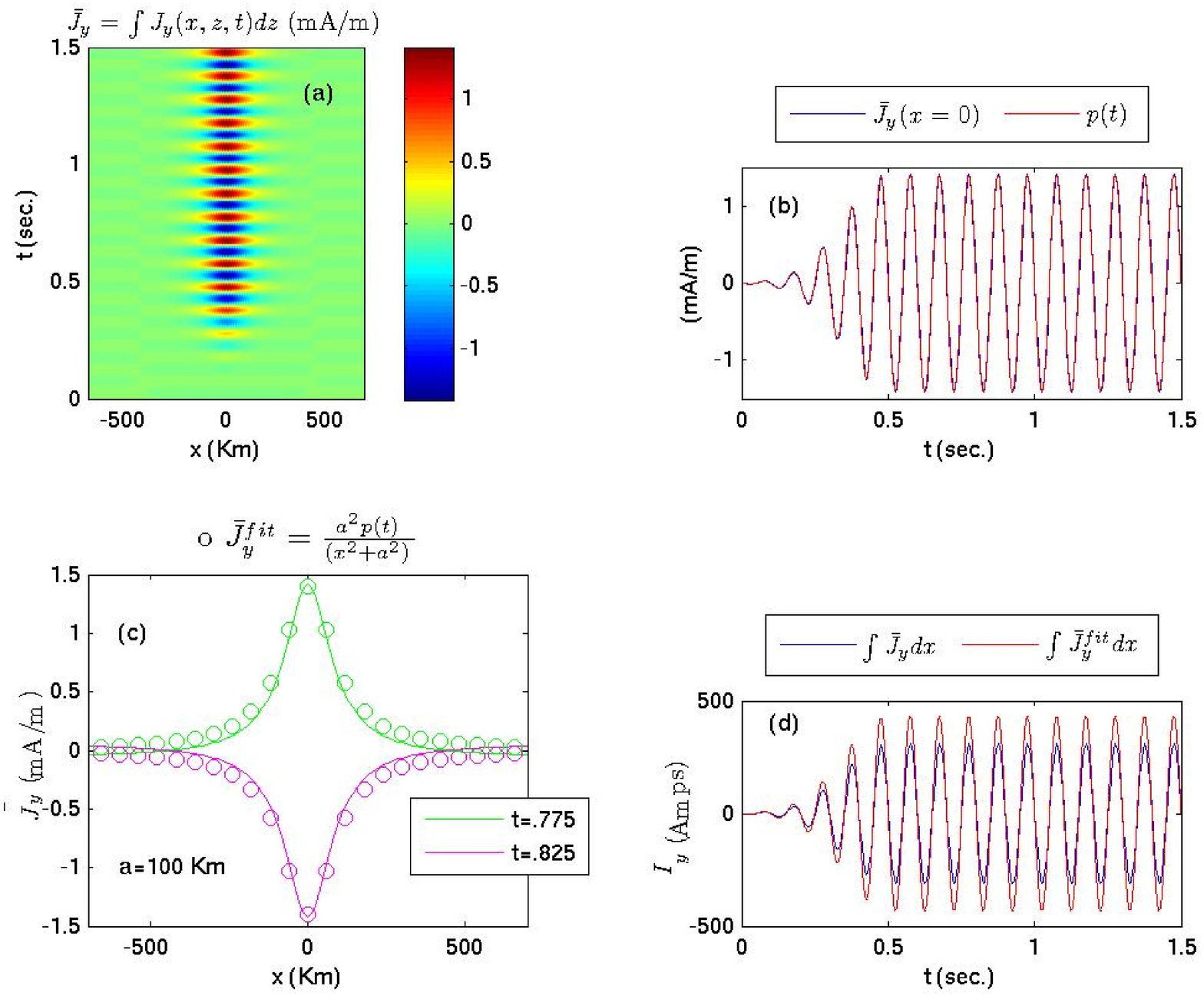}
\caption{(a) $z$-integrated current $\bar{J}_y(x,t)$, (b) time variation of $\bar{J}_y(x=0,t)$ and its functional fit $p(t)$ defined in the text, (c) $\bar{J}_y(x,t)$ as a function of $x$ and the fitting distribution $\bar{J}_y^{fit}(x,t)=a^2p(t)/(x^2+a^2)$ (represented by circles) with $a=100$ km at different times and (d) total currents along $y$ obtained by integrating $\bar{J}_y(x,t)$ and $\bar{J}_y^{fit}(x,t)$ along $x$. These results are for $\theta=0^o$, day time and continuous wave antenna field.\label{jy_fit}}.
 \end{figure}
\subsection{A current distribution fit for $J_y$}
It has been shown by \citet{park73} that for the purpose of the calculation of the electromagnetic fields produced by a current distribution $J_y\sim a/(x^2+a^2)$ above the conducting Earth, $J_y$ can be replaced by a line current along $y$ raised above its actual height by the half width ($a$) of the distribution. For this purpose, we integrate along $z$ the two dimensional current distribution $J_y$ obtained from the simulations and fit a distribution $\sim a/(x^2+a^2)$ to the integrated current. The $z$-integrated current $\bar{J}_y(x,t)=\int J_y(x,z,t) dz$ (mA/m) is shown in Fig. \ref{jy_fit}a for $\theta=0^o$, day time and continuous wave antenna field. For a given time, $\bar{J}_y$ has an $x$-profile which has its peak at $x=0$. This peak value $\bar{J}_y(x=0,t)$ oscillates in time with amplitude 1.4  mA/m in steady oscillatory state, shown in Fig .\ref{jy_fit}b which also shows the fit $p(t)$ for $\bar{J}_y(x=0,t)$. This fit $p(t)$ has the same functional form as the continuous wave antenna field except the amplitude in the steady oscillatory state, which is 1.4 mA/m for $p(t)$. Using $p(t)$, $\bar{J}_y(x,t)$ is fitted with the function,
\begin{equation}
 \bar{J}_y^{fit}(x,t)=\frac{a^2p(t)}{a^2+x^2},
\end{equation}
 where $a=100$ km. The plots of $\bar{J}_y(x,t)$ and $\bar{J}_y^{fit}(x,t)$ at two different times (when $p(t)$ is at its positive and negative peaks ) are shown in Fig. \ref{jy_fit}c. The total current $I_y$ flowing along $y$ obtained by integrating $\bar{J}_y(x,t)$ and $\bar{J}_y^{fit}(x,t)$ along $x$, shown in Fig. \ref{jy_fit}d, oscillates with amplitudes 310 Amps and 440 Amps respectively. The mismatch between the total currents is due to the slight mismatch between $\bar{J}_y(x,t)$ and $\bar{J}_y^{fit}(x,t)$ for $x>100$ km.

The $z$-integrated current shown in Fig. \ref{jy_fit} is for $\theta=0^o$, day time conditions and continuous wave antenna field. However the peak value of $p(t)$ (1.4 mA/m),  the fit $\bar{J}_y^{fit}$ and the total current $I_y$ do not change at all (except the time dependence of $p(t)$ which is different for pulsed and continuous wave antenna field) on changing the simulation parameters, viz., day and night time conductivities, pulsed and continuous wave antenna field, and values of $\theta$. This can be understood as follows. Since $\partial/\partial x << \partial/\partial z$ and $B_z$ is small as compared to $B_x$, $J_y$ can be approximated as,
\begin{equation*}
 J_y\approx\frac{1}{\mu_0}\frac{\partial B_x}{\partial z}.
\end{equation*}
 On integrating from lower boundary at $z=90$ km to the upper boundary we get,
\begin{equation*}
 \bar{J}_y(x,t)=\int J_y(x,z,t) dz =\frac{1}{\mu_0}B_x(x,z=90 km,t), 
\end{equation*}
using $B_x=0$ at the upper boundary. Due to the magnetic shielding, the value of $B_x \approx$ 1.8 nT at $x=0$ and $z=90$ km (when antenna field is at its maximum) irrespective of the simulation parameters. This gives the peak value of $\bar{J}_y(x=0,t)\approx 1.4$ mA/m, which is same as in Fig. \ref{jy_fit}b. The fit of $J_y$ along $x$ is basically related to the $x$-dependence of the $x$-component of the antenna field ($B_{x,ant}$) for a given $z$ (see Eq. \ref{bant}). Since the peak value 1.4 mA/m and $x$-dependence of the $B_{x,ant}$ remain same for all the simulation parameters, the total current $I_y$ also remains same. 
\section{Summary\label{summary}}
We have investigated the interaction of the ELF electromagnetic fields generated by an antenna placed at the Earth's surface with the E-region of the Earth's ionosphere for various parameters, viz., day and night time conductivities, pulsed and continuous wave antenna field, and different values of $\theta$. The interaction leads to the generation of the horizontal currents 
$J_x$ and $J_y$ with magnitudes up to few 100 nA/m$^2$ and vertical currents $J_z$ with magnitude up to 7 nA/m$^2$ in the E-region depending upon various simulation parameters. The associated electric fields have magnitude up to $E_x \sim 0.6$ mV/m, $E_y \sim$ 1 mV/m and $E_z \sim$ 8 mV/m. The wave penetration with a typical wavelength of the order of $\sim 10$ km of the currents and fields in to the deeper ionospheric layers up to $z\sim120$ km takes place due to the dominance of the Hall conductivity over the Pederson conductivity in the region $90$ km $<z<120$ km and penetration becomes diffusive for $z>120$ km. The scale length $\lambda$ of the vertical penetration decreases with $z$ up to $z=110$ km as $\lambda \propto 1/\sigma_H$ and $\sigma_H$ increases up to $z=110$ km. The increase in the wave speed  due to the reduced conductivities during night time leads to the deeper penetration. For continuous wave antenna field, the vertical scale length is smaller than that in the case of pulsed antenna field because of the presence of lower frequencies in the later case. The vertical scale length increases rapidly with $\theta$ up to $\theta=20^o$ and then slowly to approach an asymptotic value for large values of $\theta$. This leads to the deeper penetration with increasing value of $\theta$. The magnitudes of $J_y$ and $J_z$ near the lower boundary  decrease with $\theta$. The vertical electric field $E_z$ decreases drastically  with $\theta$, e.g., from 1.5 mV/m ($\theta=0^o$) to $10^{-4}$ mV/m ($\theta=90^o$) for pulsed antenna field and day time conditions.
The $z$-integrated current along $y$ is fitted with a current distribution $ a^2p(t)/(x^2+a^2)$ where $a=100$ km and p(t) is the peak value of the distribution. Such a current distribution can be replaced by a line current raised above its actual height by the half width ($a$) of the current distribution for the purpose of the calculation of radiation \citep{park73}. The maximum  total current along $y$ (310 Amps) remains same for various simulation parameters due to the magnetic shielding which gives $B_x \sim 1.8$ nT at $x=0$ and $z=90$ km for all the simulation parameters.

\bibliography{references}

\end{document}